\documentclass[sigconf]{acmart}

\AtBeginDocument{%
  \providecommand\BibTeX{{%
    \normalfont B\kern-0.5em{\scshape i\kern-0.25em b}\kern-0.8em\TeX}}}

\usepackage{soul}
\usepackage{xcolor}
\usepackage{amsmath,enumitem,mathtools}

\setcopyright{none}

\acmConference[CHI '23]{reCHInnecting}{April 23--28,
  2023}{Hamburg}

\begin{document}

\def\acmBooktitle#1{\gdef\@acmBooktitle{#1}}
\acmBooktitle{Proceedings of \acmConference@name
       \ifx\acmConference@name\acmConference@shortname\else
         \ (\acmConference@shortname)\fi}
         
\def\acmYear#1{\def\@acmYear{#1}}
\acmYear{\the\year}

\def\copyrightyear#1{\def\@copyrightyear{#1}}
\copyrightyear{\@acmYear}

\def\acmISBN#1{\def\@acmISBN{#1}}
\acmISBN{978-x-xxxx-xxxx-x/YY/MM}

\emergencystretch 3em  
\title{Why Combining Text and Visualization Could Improve Bayesian Reasoning: A Cognitive Load Perspective}


\author{Melanie Bancilhon}
\affiliation{%
  \institution{Washington University in St. Louis}
  \streetaddress{1 Brookings Drive}
  \city{St. Louis}
  \country{United States}}
\email{mbancilhon@wustl.edu}

\author{AJ Wright}
\affiliation{%
  \institution{Washington University in St. Louis}
  \streetaddress{1 Brookings Drive}
  \city{St. Louis}
  \country{United States}}
\email{ajwright@wustl.edu}

\author{Sunwoo Ha}
\affiliation{%
  \institution{Washington University in St. Louis}
  \streetaddress{1 Brookings Drive}
  \city{St. Louis}
  \country{United States}}
\email{sha@wustl.edu}

\author{Jordan Crouser}
\orcid{35151}
\affiliation{%
  \institution{Smith College}
  \streetaddress{1 Chapin Way}
  \city{Northampton}
  \country{United States}}
\email{jcrouser@smith.edu}

\author{Alvitta Ottley}
\affiliation{%
  \institution{Washington University in St. Louis}
  \streetaddress{1 Brookings Drive}
  \city{St. Louis}
  \country{United States}}
\email{alvitta@wustl.edu}

\renewcommand{\shortauthors}{M. Bancilhon, A. Wright, S. Ha, R. J. Crouser, and A. Ottley}

\definecolor{low}{HTML}{1b9e77}
\definecolor{high}{HTML}{d95f02}
\definecolor{single}{HTML}{386cb0}
\definecolor{dual}{HTML}{e78ac3}

\newcommand{\acro}[1]{\textsc{\MakeLowercase{#1}}}
\newcommand{\eg}[0]{e.g.,}
\newcommand{\ie}[0]{i.e.,}
\newcommand{\txt}[0]{\textit{text}}
\newcommand{\vis}[0]{\textit{vis}}
\newcommand{\vistext}[0]{\textit{vistext}}
\newcommand{\load}[0]{\textcolor{dual}{\textbf{Dual }}}
\newcommand{\Load}[0]{\textcolor{dual}{\textbf{Dual }}}
\newcommand{\noload}[0]{\textcolor{single}{\textbf{Single }}}
\newcommand{\exact}[0]{\textsc{exact }}
\newcommand{\bias}[0]{\textsc{bias }}
\newcommand{\error}[0]{\textsc{error }}
\newcommand{\ospan}[0]{\textsc{ospan }}
\newcommand{\wmc}[0]{\textsc{wmc }}
\newcommand{\tlx}[0]{\textsc{nasa-tlx}}
\newcommand{\crt}[0]{\textsc{crt}}
\newcommand{\lb}[0]{\hfill \\} 

\newcommand{\low}[0]{\textcolor{low}{\textbf{Low }}}
\newcommand{\high}[0]{\textcolor{high}{\textbf{High }}}

\newcommand{\single}[0]{\textcolor{single}{\textbf{Single }}}
\newcommand{\dual}[0]{\textcolor{dual}{\textbf{Dual }}}


\definecolor{highlighter}{HTML}{fff100}
\sethlcolor{highlighter}
\newcommand{\revision}[1]{\hl{#1}} 

\begin{abstract}
Investigations into using visualization to improve Bayesian reasoning and advance risk communication have produced mixed results, suggesting that cognitive ability might affect how users perform with different presentation formats. Our work examines the cognitive load elicited when solving Bayesian problems using icon arrays, text, and a juxtaposition of text and icon arrays. We used a three-pronged approach to capture a nuanced picture of cognitive demand and measure differences in working memory capacity, performance under divided attention using a dual-task paradigm, and subjective ratings of self-reported effort. We found that individuals with low working memory capacity made fewer errors and experienced less subjective workload when the problem contained an icon array compared to text alone, showing that visualization improves accuracy while exerting less cognitive demand. We believe these findings can considerably impact accessible risk communication, especially for individuals with low working memory capacity.
\end{abstract}





\begin{teaserfigure}
    \begin{center}
          \includegraphics[width=0.85\textwidth]{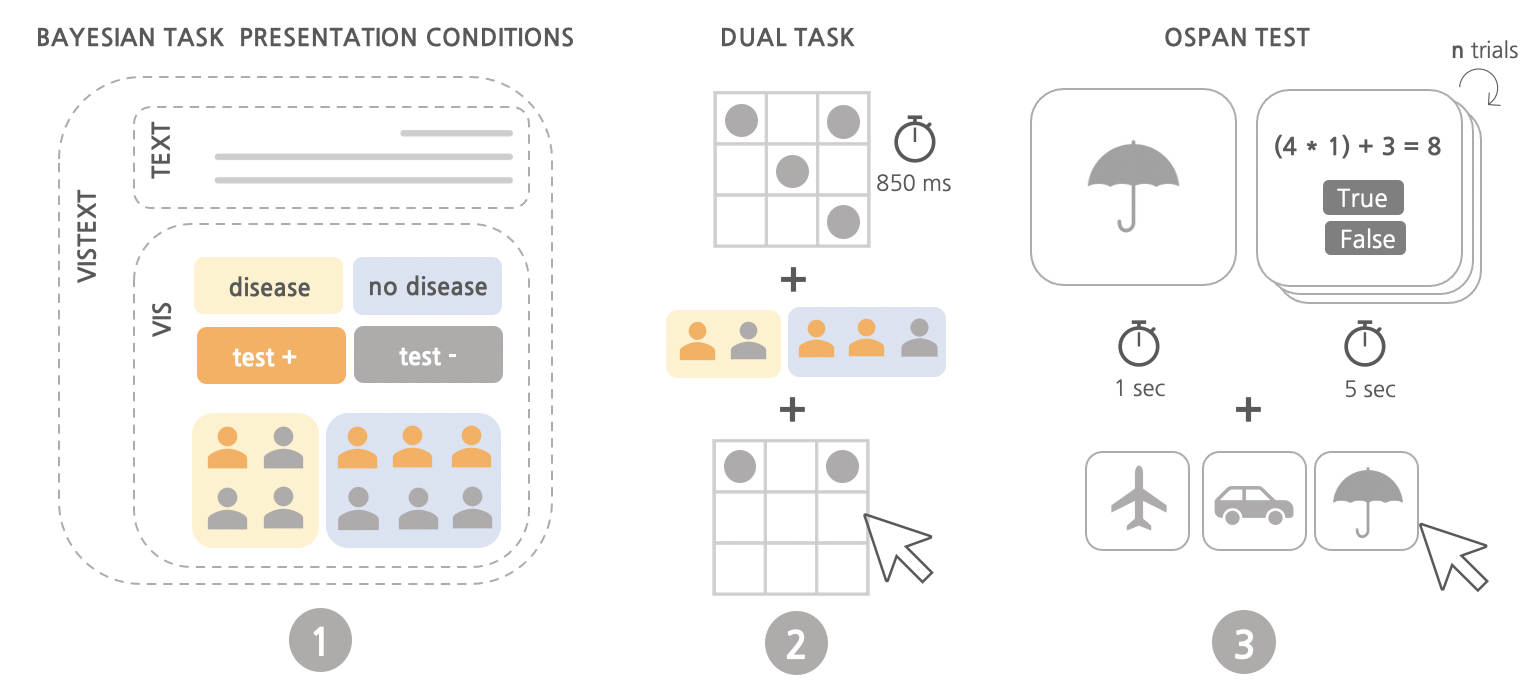}
          \caption{An illustrative overview of our experimental design, where we used cognitive load theories to study the impact of presentation format on Bayesian reasoning. 1) Users were shown a Bayesian problem via: text-only (\textit{text}), visualization-only (\vis), or combined text and visualization (\vistext) 2) We experimentally manipulated cognitive resources using a dual-task framework, asking participants to keep a 4-dot pattern in memory while completing the Bayesian task 3) We measured working memory capacity with  an operation span (\ospan) task designed by \cite{castro2019cognitive}, testing the ability to remember sequences of four, five, and six images while completing interspersed math problems. }
          \Description{An image showing an overview of the Bayesian task, the dual task, and the OSPAN test. }
          \label{fig:teaser}
    \end{center}
\end{teaserfigure}

\maketitle

\section{Introduction}

Scholars have long studied the impact of multimedia formats on comprehension and performance in various settings. In psychology, for example, studies suggest that combining a diagram and text description provides more learning benefits than showing one or the other separately (\eg  \cite{brunye2006learning,brunye2008repetition}). Similarly, in education, scholars advocate for multimedia representations over singular formats \cite{issa2011applying}. However, the guidelines are not as clear-cut for visualization, even though combining text and visualization is ubiquitous in mass media storytelling, education, and health communication.

One area in visualization research where the efficacy of combining text and visualization is fraught with uncertainty is the communication of conditional probabilities. Conditional probabilities or Bayesian reasoning is necessary to communicate crucial statistical information to a broad audience, especially in medical decision-making. In particular, health officials need to express how often a test reports that a person has a virus when they do not (false positive). Additionally, patients need to understand their chance of having the disease given a positive test (true positive) to make informed decisions about risks and potential treatment. Still, extensive research shows that understanding conditional probabilities is challenging for novices and experts alike, even with multimedia representations ~\cite{eddy1982probabilistic, kahneman2011thinking, gigerenzer1995improve,tversky2013judgment,weber2018can}.

One of the most important guidelines proposed to improve Bayesian reasoning accuracy is to show information in the form of natural frequency formats (\eg\ \textit{8 out of 10}) instead of percentages (\eg\ \textit{80\%})~\cite{lesage2013evolutionary, gigerenzer1995improve,ottley2016improving}. However, further investigations examining whether including visualization can improve Bayesian reasoning have produced mixed results. Early studies found that adding visualizations such as icon arrays to text formats can prompt faster and more accurate responses than text-only formats (\eg~\cite{brase2009pictorial}). More recent crowdsourced studies found that supplementing textual information with Euler diagrams increased accuracy only when numerical data were removed from the textual description~\cite{micallef2012assessing}, suggesting a potential conflict when presenting numbers and visualization together. Researchers have examined interaction techniques that link the text to the visualization but found no measurable benefit compared to static multimedia formats~\cite{mosca2021does}. 
Other studies have shown that spatial ability is a mediating factor for accuracy and advocate for considering individual differences in visualization evaluation~\cite{ottley2016improving}. 

The research on Bayesian reasoning presentation extends beyond the visualization community and is more expansive than the few papers we have highlighted here. Yet, despite the extensive research, our knowledge is limited, partly due to over-reliance on coarse performance measures such as reasoning accuracy. 
We propose that other factors, such as the cognitive load elicited by different presentation formats, might provide an additional window into the mechanisms underlying how people use text and visualization to support Bayesian reasoning. Cognitive load 
is a measure of the effect that a particular task has upon the user's cognitive system~\cite{paas2003cognitive}. 
It can impact user experience under various  
conditions, such as making decisions under stress or emotional burden~\cite{moran2016anxiety}, under divided attention~\cite{castel2003role, santangelo2013contribution,castro2017handheld}, or with limited mental resources~\cite{atkinson2018remember, gilchrist2008working}. 

We evaluate the cognitive load elicited by the icon array (\textit{visualization-only}), text (\textit{text-only}), and a combination of icon array and text (\textit{combined}) using three different methods:  a working memory capacity test, a dual task, and self-reported effort. We posit that measuring working memory capacity will provide insight into individual differences in users' cognitive abilities. Additionally, by burdening cognitive resources, the dual-task paradigm is a more direct method of measuring the impact of format on cognitive load and simulates real-world conditions where attention is divided. Finally, we captured perceived effort via a \tlx\ questionnaire. These three methods together provide a comprehensive view of cognitive load. 

By observing individual differences in working memory capacity, we found that individuals with low working memory made significantly fewer errors when using \textit{visualization-only} compared to \textit{text-only} formats.
Furthermore, \tlx\ scores show that users with low working memory capacity reported experiencing less temporal and physical demand using \textit{visualization-only} and \textit{combined} formats compared to text alone. Low working memory users also reported feeling less frustrated when using \textit{combined} compared to \textit{text-only}. Together, these provide supportive evidence that visualization elicits less cognitive load compared to text alone.

In summary, this paper documents the following contributions to the study of visualization-supported Bayesian reasoning:

\vspace{.5em}
\begin{enumerate}
    \setlength\itemsep{.15em}
        
    \item Using cognitive load, our findings offer a new perspective on the role of visualization for Bayesian reasoning. In particular, we found that \textbf{showing repeated information across text and visualization in combined formats could be beneficial}. We provide suggestive evidence that this enables people to select which formats better fit their mental model.
    
    \item We demonstrate that \textbf{individual differences in working memory capacity affect Bayesian reasoning} with different formats. This has implications for the use of visualization across a broad population (e.g. in medical decision-making) and adds a new dimension of complexity to the process of visualization recommendation.
    
     \item We demonstrate \textbf{how to use varying measures of cognitive load for visualization evaluation}, adding to the literature that calls for the diversification of evaluation measures by expanding beyond traditional performance metrics such as accuracy.
\end{enumerate}

\section{Background}

People are notoriously bad at reasoning with conditional probabilities~\cite{kahneman2011thinking, austin2019physician}. Consider, for example, the following scenario from~\cite{gigerenzer1995improve}:
\vspace{.5em}
\begin{quote}\label{mam-percentages}
\textit{
The probability of breast cancer in the population is 1\% for a woman aged
40 who participate in a routine screening. If the woman has breast cancer, the
probability is 80\% that she will have a positive mammography. If a woman
does not have breast cancer, the probability is 9.5\% that she will also have a
positive mammography. A woman in this age group had a positive
mammography in a routine screening. \newline \newline
What is the probability that she actually has breast cancer?\hspace{.01\linewidth}\rule{.1\linewidth}{0.05mm}}
\end{quote}

\noindent
According to Bayes' theorem, 
 \begin{equation}
   P(H|D) = \frac{P(D|H) \times P(H)} {P(D)} 
 \end{equation}

\noindent
where, in our scenario, $D$ is the positive mammography and $H$ is the hypothesis that the woman in question has breast cancer.
It is common for people, including experts, to be subject to \textit{base-rate neglect}, ignoring the base rate $P(H)$ when reasoning about the true positive rate~\cite{kahneman2011thinking}. For decades, there have been efforts across various fields to devise ways to improve Bayesian reasoning by mitigating the base rate fallacy.

Several studies have shown that frequency formats (\eg\ 8 out of 10 instead of 80\%) can facilitate Bayesian reasoning and significantly improve accuracy~\cite{lesage2013evolutionary,gigerenzer1994distinction,gigerenzer1995improve,cosmides1996humans,brase2008frequency}. 
Additionally, many researchers have investigated the effect of visualization on Bayesian reasoning (\eg~\cite{cole1989understanding,cole1989graphic,khan2018interactive,spiegelhalter2011visualizing, khan2015benefits}), with the most prevalent designs being Euler diagrams~\cite{brase2009pictorial,kellen2007facilitating,khan2015benefits,micallef2012assessing} and frequency grids or icon arrays ~\cite{bocherer2019improve,garcia2013visual,ottley2012visually,sedlmeier2001teaching,kellen2007facilitating,khan2015benefits,micallef2012assessing,tsai2011interactive}. These designs represent two dominant theories behind Bayesian facilitation. Euler diagrams align with the \textit{nested set theory}. They are useful to help the viewer reason about how subsets relate to each other\cite{sedlmeier2001teaching,tversky1981evidential,lesage2013evolutionary,barbey2007base}, while icon arrays, showing natural frequencies (i.e., 8 out of 10), align with the \textit{ecological rationality framework} positing based on evolutionary theories that humans are better at reasoning with countable objects \cite{cosmides1996humans,gigerenzer1994distinction}. Our work uses icon arrays because of their popularity and the well-documented success of natural frequency formats for Bayesian reasoning (\eg~\cite{gigerenzer1995improve,cosmides1996humans,brase2008frequency,ottley2016improving,micallef2012assessing}).

To investigate the potential benefit of visualization in Bayesian reasoning, researchers have typically compared responses to Bayesian problems presented in text format to formats that combine visualization and text. However, these studies have produced mixed findings.  
For example, Micallef et al.~\cite{micallef2012assessing} found no measurable difference in accuracy between text alone and a combination of text and visualization. Still, their follow-up study demonstrated that removing the numbers from the text significantly improved Bayesian accuracy. Ottley et al.~\cite{ottley2016improving,ottley2019curious} replicated this first study result and found no overall reliable differences in accuracy between the text alone versus a combined format. However, they found that participants with high spatial ability performed reliably better with visualization alone compared to text alone~\cite{ottley2016improving}. In another study, Ottley et al.~\cite{ottley2019curious} used eye-tracking to examine how people extract information from text-only, visual, and combined formats in Bayesian reasoning problems. They found that users easily identify information with visualization but extract information more easily from the text. Additionally, their analysis found no differences in how the study participants used each format when they saw the combined presentation. Finally, Mosca et al.~\cite{mosca2021does} investigated the effect of linking the text and visualization via interaction. They found that adding interaction did not improve accuracy in Bayesian reasoning compared to static formats.

We posit that the outstanding questions on whether visual designs can improve Bayesian reasoning could be due to a lack of understanding of underlying cognitive mechanisms. Investigations by Lesage et al.~\cite{lesage2013evolutionary} showed that performance in Bayesian reasoning is reliant upon available mental resources, regardless of presentation format. Although visualization researchers often seek to improve speed and accuracy measures, we know little about the impact of visualization on cognitive load. Moreover, speed and accuracy do not always correlate with cognitive load when reasoning about visualizations~\cite{huang2009measuring, riveiro2014effects}. Thus, there is a need to understand the processes that govern Bayesian reasoning with different presentation formats. In this paper, we expand the evaluation of Bayesian communication techniques by measuring cognitive load through individual differences in working memory capacity, a dual-task paradigm, and perceived cognitive load. We aim to develop a more nuanced understanding of the potential effect of presentation formats on Bayesian  facilitation and provide more comprehensive visualization design guidelines. 

\subsection{Measuring Cognitive Load}

Working memory consists of multiple components that can store a limited amount of information for a limited amount of time and is an essential resource in the reasoning process~\cite{cowan1997role}. Cognitive load, typically defined as the amount of working memory required to process a task, is an important usability factor that indicates how easy or how hard it is to process information~\cite{paas2003cognitive}. There exist numerous techniques to measure cognitive load, including self-reported measures (e.g. \tlx), performance-based measures (e.g. dual-task paradigm, operation span tests) and physiological measures (e.g. pupillometry, f\acro{NIRS})~\cite{kennedy2000glucose, mulder1992measurement, brunken2003direct,lipp2004attentional, huang2009measuring, padilla2018,de1996measurement,paas2003cognitive}. Several researchers have leveraged these techniques to investigate the effect of visualization design on cognitive load~\cite{tintarev2016effects,borgo2012empirical,peck2013using,anderson2011user,zhu2010visualization, castro2019cognitive}, sometimes reexamining long-standing beliefs. For example, Matthews et al. highlight the importance of using several methods to cross-examine the effect of workload~\cite{matthews2015psychometrics}. In their work, Borgo et al. challenged traditional notions about chart junk and showed using a dual-task paradigm that visual embellishments do not prompt higher cognitive load compared to other visualizations~\cite{borgo2012empirical}.
Peck et al. used f\acro{NIRS} as well as \tlx\ to evaluate visualization interfaces and found no difference in the cognitive load elicited by bar graphs and pie charts, contrarily to popular belief~\cite{peck2013using}.

While physiological measures have proven to be effective techniques for measuring cognitive load, their high intrusiveness makes them unsuitable for real-life implementation~\cite{khawaja2014measuring}. Other measures are more accessible, facilitate longitudinal studies, and allow us to survey a diverse population. 
In our work, we chose to investigate the effect of presentation formats on cognitive load for Bayesian reasoning using three different methods: an operation task to observe individual differences in working memory capacity, a dual-task paradigm, and self-reported scores through a  \tlx\ questionnaire.

\subsubsection{Individual Differences Approach to Cognitive Load}

Individual differences can impact how we reason with different formats (see~\cite{liu2020survey} for a comprehensive review of individual differences in visualization), and there is strong evidence that cognitive traits can influence statistical reasoning~\cite{ottley2016improving,mosca2021does,lesage2013evolutionary, yin2020bayesian}. Some researchers showed evidence that when information was presented in the form of natural frequencies, participants with high working memory capacity performed significantly better than participants with low working memory capacity~\cite{lesage2013evolutionary, yin2020bayesian}. Castro et al. \cite{castro2021} have shown that visualization designs can elicit different levels of cognitive load when reasoning about uncertainty visualizations.

A test that has shown high correlations with measures of working memory capacity is the Cognitive Reflection Test (\acro{CRT})\cite{lesage2013evolutionary}. The \acro{CRT} test measures one's ability to overcome heuristics and biases and trigger analytical thinking~\cite{lesage2013evolutionary}. A more direct way of measuring individual differences in working memory capacity is by using an operation span task (\ospan)~\cite{castro2019cognitive}. In a typical \ospan\ task, participants must simultaneously try to remember presented words in their correct order while solving simple math equations sequentially. In this paper, we use Castro et al.'s adapted online \ospan\ test to measure working memory capacity~\cite{castro2021} \footnote{Link to \ospan test used in this work (developed by \cite{castro2021}): \url{https://bit.ly/2QHErIv}}. To complement this method, we use a dual-task paradigm, which according to Lesage et al.~\cite{lesage2013evolutionary}, can be used to infer a causal role for cognitive resources in the performance of Bayesian reasoning tasks. 

\subsubsection{Dual-Task Paradigm}
Although it has not been prominently featured in visualization research, the \textit{dual-task methodology} is an effective way to assess the dependency of a task on cognitive resources and has been used to evaluate workload in psychology for decades ~\cite{pashler2016attention,chiles1982workload}. In a dual-task paradigm, the user conducts two tasks simultaneously, a primary task and a secondary task. This creates \textit{divided attention} and increases cognitive load, producing a decline in performance compared to the primary task alone. This decline is  often referred to as the \textit{dual-task cost}~\cite{padilla2019toward}, which can be used to infer the cognitive load elicited by the task.

Several researchers have investigated the impact of formats on cognitive load using a dual-task paradigm~\cite{borgo2012empirical, castro2017handheld,tintarev2016effects}, one reason being that it is helpful to simulate real-life conditions where attention is often divided~\cite{castro2017handheld, castel2003role,santangelo2013contribution}. Castro et al.~\cite{castro2017handheld} have used a dual-task method to investigate how display dimensions and screen size of mobile devices influence attention. In their study, participants controlled the movements of a blue ball by tilting the mobile device on displays of different sizes (primary task) while performing a change detection task which consisted of vocally reporting which of 4 arrows changed directions on a fixed display (secondary task). Using this methodology, they found that larger displays are more mentally demanding under divided attention. Tintarev et al.~\cite{tintarev2016effects} investigated the effect of presentational choices for \textit{planning} on cognitive load using a dual-task paradigm. Participants had to keep information about a list of words in memory while answering some questions about a plan, then had to recall the list of words in the correct order. The authors found no reliable differences in performance across different formats of the plan. 

In our work, we quantify differences in elicited cognitive load across presentation formats using a dual-task methodology inspired by \cite{lesage2013evolutionary}, consisting of remembering a pattern of four dots on a grid while conducting the primary task.

\begin{figure*}[t]
    \begin{center}
          \includegraphics[width=0.85\textwidth]{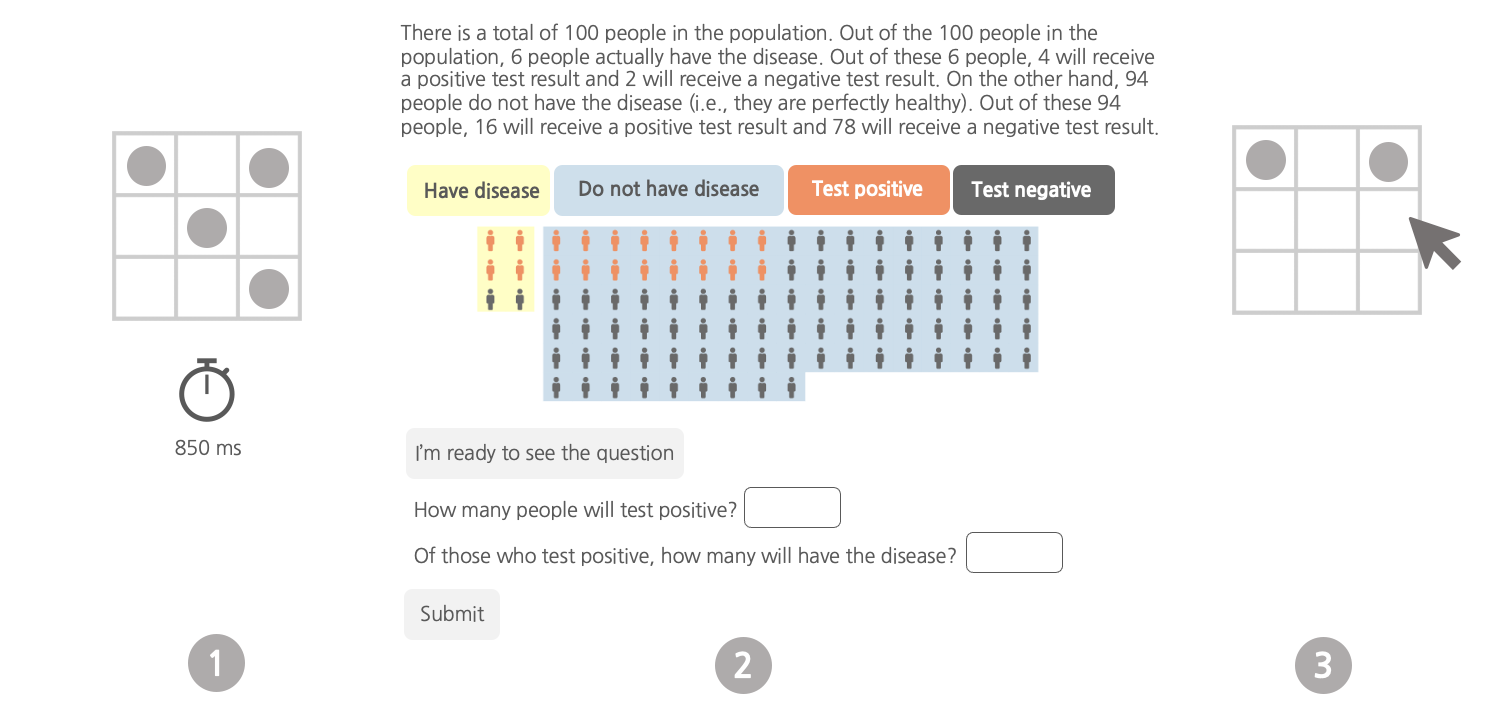}
          \caption{An overview of the Bayesian survey for  the \dual\ condition with the \vistext\ format 1) Users were shown for 850 ms a pattern consisting of four dots on a 3x3 grid that they were asked to memorize 2) This is an example of the Bayesian task for the \vistext\ condition. Users were asked to read the problem and then press a button when they were ready to answer questions 3) Once users submitted their answers to the Bayesian questions, they were asked to replicate the dot pattern on an empty 3x3 grid. }
          \Description{An image of the steps involved in the dual-task paradigm.}
          \label{fig:experimentalDesign}
    \end{center}
\end{figure*}

\section{Research Goals}

We designed two complementary studies to investigate whether cognitive load can shed light on the conflicting and sometimes puzzling findings around Bayesian reasoning and visualization. 

These findings collectively point to a potential relationship between cognitive resources and Bayesian facilitation --- adding visualization and interaction to an already cognitively challenging task might not produce the desired effects. There is a gap in our understanding of how cognitive load affects Bayesian reasoning across different formats. Motivated by this, the current work focuses on examining the potential differences in cognitive load elicited by visualization-only, text-only, and a combination of text and visualization format in the context of Bayesian reasoning. We use the icon array for our visualization condition because it is prominently used to communicate Bayesian information, especially in the context of medical risk, supporting ecological validity. 

When considering options for the experiment design, we weighed trade-offs between (1) controlling the framing and learning effects, (2) minimizing noise from individual variability, and (3) minimizing the overall length of the study.
Unfortunately, no single experiment strikes the perfect balance. Thus, we present the results of two controlled user experiments. The first adopts a between-subject, 3 (\textit{presentation format}) $\times$ 2 (\textit{load condition}), experiment design to mitigate the learning effects that a within-subject study would introduce. The second utilizes a mixed design, with 3 (\textit{presentation format}) between-subject and $\times$ 2 (\textit{load condition}) within-subject protocol to better control for individual variability. Together, they tell a cohesive story about the relationship between cognitive load, Bayesian reasoning, and visualization.

\section{Experiment 1: Between-Subject Study Design}
\label{sec:method}

We assigned each participant randomly to one of three presentation conditions --- icon array (\textit{vis}), text (\textit{text}), and a combination of icon array and text (\textit{vistext}) --- making the comparison of presentation between subject. We also assigned each user randomly to either a \single or \dual task, making the comparison of these tasks also between subjects. We chose a between-subject design to keep the Bayesian problem consistent across all conditions. Prior work has shown that different Bayesian scenarios can lead to different levels of accuracy \cite{micallef2012assessing} \footnote{Link to Experiment 1 surveys, data, and analyses:~\url{https://bit.ly/3BFwokx}}.

\subsection{Presentation Formats and Bayesian Task}
\label{sec:bayes-task}

We replicated Mosca et al.'s~\cite{mosca2021does} grouped icon array design, which had the highest accuracy among their tested visualization formats. The authors designed the icon array according to Bertin's\cite{bertin1983semiology} guidelines, where background color was used to differentiate between members of the population who \acro{HAVE DISEASE} versus \acro{DO NOT HAVE DISEASE} and icon color was used to differentiate between members of the population who \acro{TEST POSITIVE} versus \acro{TEST NEGATIVE}. Participants in our the \txt\ condition saw the same data in textual format, and those in the \vistext\ condition saw both the textual format and the icon array, vertically stacked.

We showed participants data about the prevalence of a disease in a population, as well as the test results in the form of either \vis, \textit{text} or \textit{vistext}. We asked them to estimate i) the number of people who will test positive and ii) of those people, how many actually have the disease. This technique of prompting the user for the positive count followed by the true positive count is called \textit{probing}. \textit{Probing} is a valid technique that evaluates Bayesian comprehension independently of mathematical skills through the retrieval of nested data (using the words "of those"). It has been shown to elicit more accurate responses compared to non-probed questions \cite{cosmides1996humans, brase2009pictorial, ottley2016improving}.

\subsection{Load Conditions and Dual-Task Methodology}
Participants either saw the Bayesian probability estimation task alone or along with a secondary task. 
Participants who were randomly assigned the \dual condition were shown a pattern consisting of four dots on a grid for 850ms and were asked to complete the Bayesian Probability Estimation task while keeping the pattern in memory. Participants were then asked to reproduce the dot pattern as accurately as possible by selecting the appropriate cells on an empty grid. \autoref{fig:teaser} illustrates the dual task setup, inspired by Lesage et al.'s~\cite{lesage2013evolutionary} study of text-only formats and originally developed by Bethell et al.~\cite{bethell1988mental}. This task is appropriate as it taxes visuospatial working memory, which would possibly interfere with the primary task and cause the desired increase in cognitive load.


\subsection{Measures of Abilities and Surveys}

The survey also contained a NASA-TLX questionnaire, a spatial ability test, and a Cognitive Reflection Test (CRT). Participants then completed a working memory capacity questionnaire from \cite{castro2019cognitive}.

\textbf{NASA-TLX. } We used the \tlx~\cite{hart1988development,hart2006nasa} to examine participants' subjective workload. Participants reported on the workload they believed the Bayesian task elicited on six subscales: mental demand, temporal demand, frustration, physical demand, performance, and effort. 

\textbf{Working Memory Capacity Test (OSPAN). }  
We asked participants to remember a series of objects sequentially while answering simple True or False math problems. The test consisted of 6 sequences of 4-, 5- or 6- spans, shown two times each in a randomized order (the term \textit{n}-span refers to the sequence occurring \textit{n} times). In each span, participants were shown an image for 1 second and were asked to keep it in memory while answering a simple math question in under 5 seconds. This sequence is repeated \textit{n} number of times and at the end of the span, participants have to recall the images shown in the correct order. This version of the OSPAN has been designed by Castro et al.~\cite{castro2017handheld}.

\textbf{Cognitive Reflection Test. }
The Cognitive Reflection Test (CRT) has been shown to be a valid measure of cognitive load~\cite{lesage2013evolutionary}. In our work, we use a version of the CRT test that contains 3 questions. It tests for the ability to switch from Type 1 (intuitive) to Type 2 (strategic) reasoning. Since the latter requires using working memory~\cite{kahneman2011thinking}, researchers posit that someone who is able to perform the switch has a high working memory capacity~\cite{padilla2018}. 

\textbf{Spatial Ability Test. } A spatial ability test measures an individual's capacity to process visual and spatial information. In this study, we used the paper folding test (VZ-2) from Ekstrom, French, and Hardon~\cite{ekstrom1976manual} consisting of two sessions of 3 minutes and 10 questions each. This test has been used as a standard technique to evaluate Bayesian reasoning performance across spatial ability in other studies~\cite{kellen2007facilitating, micallef2012assessing}.

\textbf{VisText Usage Report. } We asked participants in the \vistext\ condition what percentage of the visualization and the text they utilized to answer the Bayesian questions. They reported their preferred method by selecting the appropriate value on a scale ranging from \textit{only text} to \textit{only visualization}

\subsection{Hypotheses}
    \begin{enumerate}
        \item [ \textbf{H1}] We hypothesize that performance on the Bayesian reasoning task depends on available cognitive resources. Therefore, the \single condition will result in more accurate reasoning than the \dual condition. 
        
    \item [\textbf{H2}] Since available cognitive resources are mediated by working memory capacity, we expect that individuals with high working memory capacity will be more accurate than their low working memory counterparts, especially in the \dual condition.

    \item [\textbf{H3}] Prior work that examined the impact of text-only, icon array, and the juxtaposition of text and icon array on Bayesian reasoning found no significant difference in accuracy between the three presentation formats \cite{micallef2012assessing,ottley2016improving}. Therefore, we anticipate no significant difference in Bayesian reasoning accuracy across \textit{vis}, \textit{text}, and \textit{vistext} in the \single condition. 

    \end{enumerate}

The detailed analysis for pre-registered hypotheses H4a - H4d can be found in the supplementary material.

\subsection{Participants}
\label{ssec:participantsExp1}
We recruited users via Amazon's Mechanical Turk that were from the United States, were English-speaking, and had a HIT acceptance rate of 100\%. 

\textbf{Payment. }  All participants were paid in accordance with minimum wage laws, on average receiving \$4.84 and taking 25.2 minutes to complete both surveys. In the Bayesian task, participants won a bonus of \$0.50 for each question answered. Participants in the \dual condition were assigned an additional task, increasing the amount of time spent on the task, and thus received an additional \$0.25 for each dot correctly remembered (i.e. up to \$1 additional bonus compared to the \textit{single} condition). The allocated bonus per dot remembered also served as an incentivization to remember the pattern.

We conducted a statistical power analysis using the software G*Power on a mixed ANOVA and determined that the target sample size needed for a statistical power of 95\% is 251.
We recruited 450 participants due to the typically high number of exclusions in Mechanical Turk studies. Users were asked to complete two separate surveys: the Bayesian survey and the OSPAN survey. Our pre-registered exclusion criteria \footnote{Link to Experiment 1 pre-registration: \url{https://bit.ly/3xtC1zX}}, determined based on prior work, required that users i) take the surveys only once ii) complete both surveys iii) score above chance in the math portion of the OSPAN test iv) score above 10\% in the memory portion of the OSPAN test v) score over 2 standard deviations from the mean in the dot pattern task. After excluding data that did not fit the exclusion criteria, 316 participants remained. After preliminary data analysis, we noticed some additional fraudulent and invalid responses that we had not anticipated prior to the pre-registration. We decided to exclude users who entered more than 4 dots in the dot pattern recall test, thus biasing their odds of getting the correct pattern (n=13). We also excluded participants whose answers to the Bayesian questions were less than or equal to 0 (n=4), which shows a lack of attention and leads to an invalid \error value upon data processing (see section 4.6). We conducted a post hoc sensitivity analysis that showed that the addition of the two exclusion criteria did not affect the study results (see supplementary material). After these non-pre-registered exclusions, 299 participants remained, of which 104 were assigned \txt, 100 were assigned \vis, and 95 saw the \textit{vistext} (129 in the \dual condition and 170 in the \single).

\subsection{Data Collection}

\noindent
The independent variables for this experiment are:
\vspace{.5em}
\begin{itemize}
    \item \textbf{3 presentation formats:} \{ \textit{text}, \vis, \vistext \}
    \item \textbf{2 load conditions:} \{ \single,  \dual \}
\end{itemize}

\noindent
To measure Bayesian performance we calculated the true positive rate from the participant's response as described in \autoref{sec:bayes-task}. Our dependent variables were:
\vspace{.5em}
\begin{itemize}
    \item \textbf{\exact} $\in \{0, 1\}$, binary value for whether the response was exact.
    \item \textbf{\bias} is the $log_{10}$ ratio of the response and the ground truth.
     \item \textbf{\error} is the absolute value of bias. 
\end{itemize}

While \exact evaluates verbatim comprehension, \bias and \error are proxies for gist (approximate) comprehension, which is more prominently used for reasoning and decision-making \cite{ETNEL2020104005,reyna2008theory,reyna2004people,padilla2018,bancilhon2020did}. 


\noindent
The covariates and other computed measures were:
\vspace{.5em}
\begin{itemize}
    \item \textbf{\ospan} $\in [0...30]$, measures general cognitive capacity.
    
    \item \textbf{\wmc} $\in \{low, high\}$, based on a median split of \ospan\ scores.
    
    \item \textbf{\tlx}  $\in [0...20]$, measures combined subjective workload. 
    
    \item \textbf{Spatial Score}  $\in [-4...20]$, is the spatial ability test score. 
    
    \item \textbf{Spatial Level} $\in \{low, high\}$, from a median split of spatial scores.
    
    \item \textbf{\crt} $\in \{0,1,2,3\}$, is the cognitive reflection test score. 
    
    \item \textbf{Text-Vis Usage} $\in [1...20]$, maps 0 to using primarily text and 20 to mostly visualization for those in the  \vistext\ condition.

\end{itemize}

\begin{figure*}[h]
    \begin{center}
          \includegraphics[width=1\textwidth]{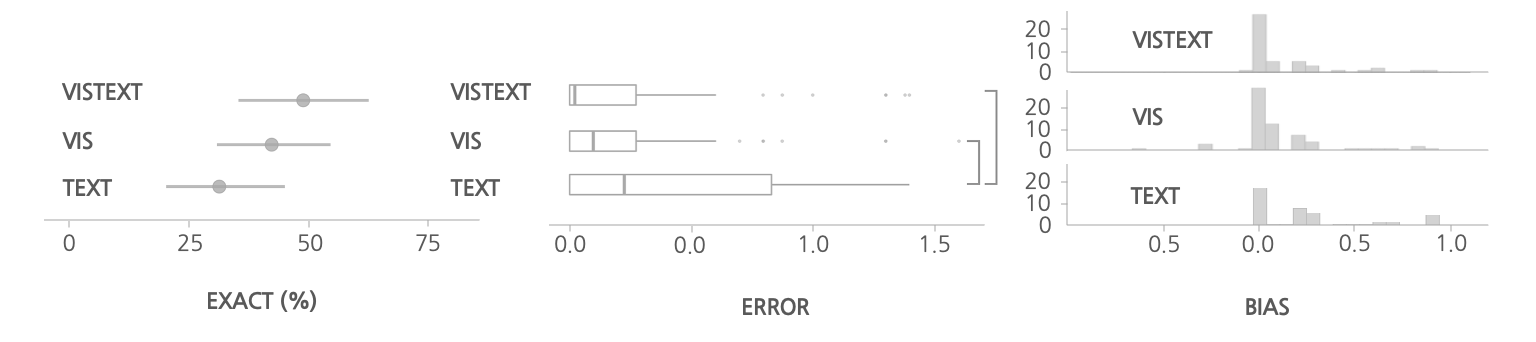}
          \caption{\single\ task \exact (95\% CI), \error\ and \bias across presentation formats. ] indicates a significant difference between the two formats $(\alpha=0.0167)$. We found significant differences in \error between \vis  and \textit{text}, and \textit{vistext} and \textit{text}.}
          \Description{A graph of the proportion of correct answers for each presentation format and the error for each presentation format.}
          \label{fig:exp1Baseline}
    \end{center}
\end{figure*}

\begin{table}[b]
  \caption{Experiment 1 condition-wise dropout rates for the Bayesian Task}
  \label{tab:attritionExp1}
  \begin{tabular}{p{6cm}c}
    \toprule
    Load condition & Dropout rates\\
    \midrule
    \single: Participants conducted the Bayesian Task & 3.98\% \\
    \dual: Participants conducted the Bayesian Task and a secondary recall task & 7.18\% \\
    \bottomrule
  \end{tabular}
\end{table}
    
\subsection{\textbf{Attrition Analysis}}
\label{ssec:exp1attrition}
\lb{}
There has been a growing body of work about the issue of high attrition rate in online studies  \cite{kraut2004psychological,zhou2016pitfall,reips2000web}. According to research by Zhou et al.~\cite{zhou2016pitfall}, studies that are cognitively taxing should be concerned if dropout rates are 20\% or above. The authors also highlight the importance of checking for selective attrition by making sure the dropout rates are not significantly different across experimental conditions. To provide transparency and encourage practices that improve internal validity, we conducted an attrition rate analysis as recommended by Zhou et al. \cite{zhou2016pitfall}.

Our experiment consists of two surveys, a Bayesian Task implemented by the authors and an OSPAN test from \cite{castro2019cognitive} on Qualtrics. We conducted an attrition analysis for the Bayesian task, where participants were assigned either a single task (\single) or a dual task (\dual). We adapted our methodology from Zhou et al. \cite{zhou2016pitfall} and only took into account participants who consented to the study and discarded fraudulent responses where participants took the survey more than once by using their recorded IP addresses. Table \ref{tab:attritionExp1} shows the condition-wise dropout rates, computed according to \cite{zhou2016pitfall} by dividing the number of participants who were assigned to a given condition and completed the entirety of their task \footnote{our server recorded an end-of-experiment timestamp when a participant completed the entire survey} by the number of those who were assigned to the same condition who at least gave their consent and only took the task once. We observe low dropout rates for both conditions that have no significant difference $(\chi^2(2) = 1.88, p = 0.1704, d=0.1406)$.

\begin{figure*}[h]
         \centering
         \includegraphics[width=0.75\linewidth]{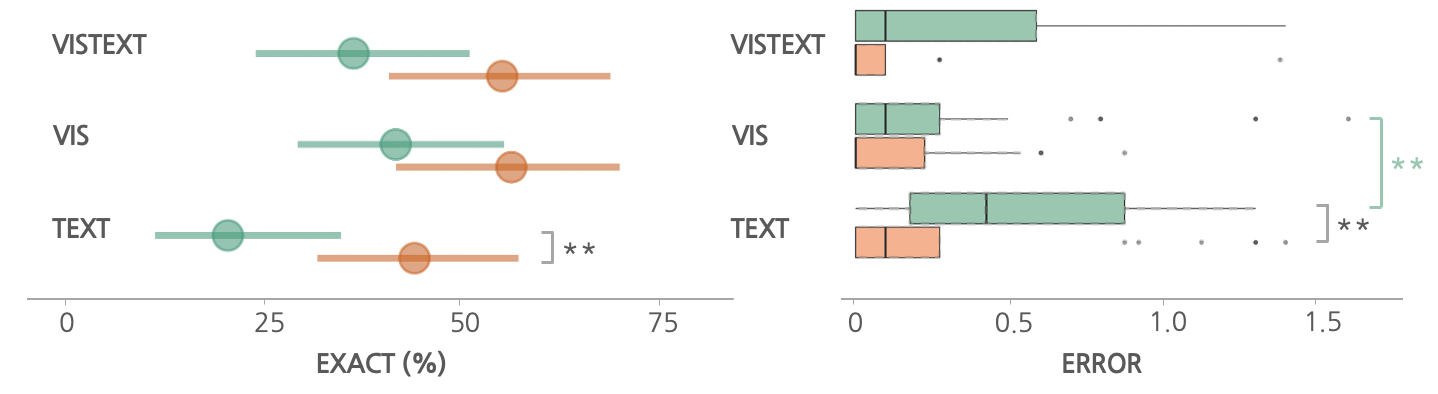}
         \label{fig:error_wmc_noload}
         \vspace{-1.5em}
        \caption{\single\ task \exact\ (95\% CI) and \error\ across presentation format and working memory group. ** indicates a significant difference between groups. We found significant differences in \exact\ and \error\ between \low\ and \high\ working memory groups $(\alpha=0.05)$ in the \text\ condition. Among \low working memory capacity individuals, \error\ was significantly higher in \textit{text} compared to \textit{vis} $(\alpha=0.0167)$}
        \Description{A graph of the proportion of correct answers for each presentation format and the error for each presentation format, separated by working memory capacity (low and high)}
        \label{fig:e1_single_exact_error}
\end{figure*}

\subsection{Findings}

Out of 299 participants, 104 were assigned \txt, 100 were assigned \vis, and 95 saw the \vistext. Each participant completed a single Bayesian problem depicting the \textit{disease} scenario in \autoref{sec:bayes-task}. Further, 170 were assigned to the single task (\noload) condition and 129 were assigned the dual-task condition with an added load (\dual).

\subsubsection{\textbf{Single Task: Establishing a baseline}} 

\lb{}
We begin our analysis by inspecting how participants performed under the single task (\noload) condition and testing whether format influence performance. The existing literature has produced mixed results on the effect of visualization on reasoning accuracy~\cite{ottley2016improving,micallef2012assessing,ottley2019curious}, and our \textbf{H3} posits no significant difference in Bayesian reasoning accuracy.

\paragraph{\bias} We conducted an exploratory analysis by examining how much participants' responses deviated from the \exact\ answer and the effect of format on their discrepancy. We observe an overall median \bias\ of .10 for the single task condition with varying median \bias\ of 0.22 for \txt, 0.00 for \vis, and 0.00 for \vistext. 
From \autoref{fig:exp1Baseline}, we can observe that participants' \bias\ are not normally distributed. Thus, we use non-parametric tests for our analysis.
Additionally, participants in the \vis\ and \vistext\ conditions were marginally more likely to produce the \exact\ answer (\bias\ $= 0$) than those who used \txt. When we ran a 3-way Kruskal-Wallis test with presentation format as a between-subject factor we found a significant difference in \bias\ across the three conditions ($H(2) = 12.87, p = .0016, \eta^2(H)= 0.065$). Follow-up Mann-Whitney Wilcoxon tests with an adjusted alpha $\alpha=0.0167$ revealed significant differences in \bias\ between \vis\ and \textit{text} ($W = 2379.5, p= 0.0006, \eta^2(H)= 0.092)$ as well as \vistext\ and \textit{text} $(W = 1772.5, p = 0.0087, \eta^2(H)= 0.057)$.

\paragraph{\textsc{exact}} 
We examined our first measure of accuracy, \exact, to investigate whether the presentation format influences the proportion of correct answers. Overall, 40.9\% of participants correctly answered both Bayesian questions for the single task condition, with \txt, \vis, and \vistext\ yielding 31.5\%, 43.08\%, and 49.02\% exact answers respectively. Our omnibus proportion z-test shows no significant effect of presentation format on accuracy $(\chi^2(2) = 3.4899, p = .1795)$. Thus, \textit{the proportion of successful exact reasoning did not depend on presentation format.}

\paragraph{\error} For a more fine-grained measure of accuracy, we examined \error\ to assess how far participants' responses deviated from the \exact\ answer and whether presentation format mediated this effect.
The median \error\ was 0.097 overall and 0.097 in the \vis\ condition, 0.22 in the \textit{text} condition, and 0.021 \vistext\ condition. A Kruskal-Wallis non-parametric test revealed significant differences in \error\ between conditions ($H(2) = 8.43, p = 0.0148, \eta^2(H)= 0.037$).
Post-hoc Mann-Whitney Wilcoxon tests with an adjusted alpha $(\alpha= .0167)$ revealed significant differences between \vis\ \& \textit{text} ($W = 2227.5, p = .0093, \eta^2(H)= 0.0047)$ and \vistext\ \& \textit{text} ($W = 1738.5, p = .0164, \eta^2(H)= 0.046$). We found no significant difference between \vis\ and \vistext\ ($W = 1610.5, p = 0.675)$. These findings suggest that \textit{reasoning with text-only led to significantly higher errors compared to other formats}.

Altogether, these findings show evidence that presentation format can impact reasoning errors. However, the observed effects were small and there was no significant impact on \exact\ response rates. Thus, \textbf{our results only partially support H3}. More specifically, they suggest that visualization, even when combined with text, can have benefits on Bayesian accuracy compared to text alone. It is noteworthy that these results also partially contradict the visualization literature that compared Bayesian formats. On one hand, our findings are similar to Ottley et al.~\cite{ottley2016improving,ottley2019curious} who found no difference in \exact\ between \textit{text}, \vis, and \vistext, but did not examine \error. On the other hand, our results differ from  Micallef et al.~\cite{micallef2012assessing} who examined \textit{text}\ and \vistext\ and found no measurable effect of these formats on \exact\ or \error. However, the discrepancies between our results and prior work could be attributed to the type of visualization used and differences in the experiment design. 

\begin{figure*}[h]
     \centering
         \centering
         \includegraphics[width=0.75\linewidth]{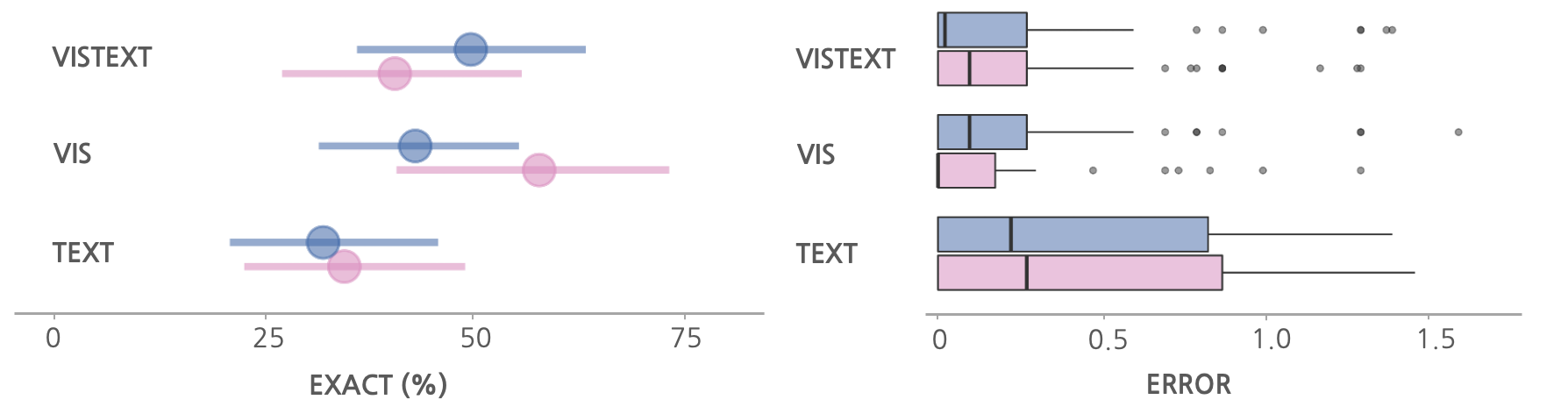}
         \label{fig:error_wmc_noldsoad}
         \vspace{-1.5em}
        \caption{\exact\ (95\% CI) and \error\ for \single and \dual task}
        \Description{A graph of the proportion of correct answers for each presentation format and the error for each presentation format, separated by single and dual task}
        \label{fig:exp1_exact}
\end{figure*}

\subsubsection{\textbf{Individual Differences in Working Memory Capacity}}
\lb{}
\label{sssec:workingmemFindingsExp1}
A primary goal of this project is to examine whether cognitive resources can explain Bayesian reasoning results. Specifically, with \textbf{H2}, we hypothesized that accuracy in Bayesian reasoning will depend on available cognitive resources. To this end, we examine the effect of working memory capacity on accuracy in the \single\ task.

To examine whether working memory mediates accuracy in participants' \exact\ and \error\ measures. We first performed a binary logistic regression to test for the effect of \ospan\ on \exact and found that correctly answering the Bayesian questions is 1.49 times more likely to occur for every 5-point increase in the working memory test (95\% CI [.04, .12]).
Analyzing \error, a generalized linear model also revealed a significant impact of \ospan\ on \error\ $(t(169)=-3.326, p =0.00108)$. Thus, \textit{the higher their working memory capacity, the more accurate participants were in their answers}. 

Following prior work~\cite{castro2019cognitive}, we split participants into \low\ and \high\ working memory groups based on a median split of their \ospan\ scores. \autoref{fig:e1_single_exact_error} summarizes the accuracy of each working memory group across presentation formats, showing their respective proportions of \exact\ answers  and \error distribution. Overall, in the \single\ task, 52.29\% of those in the \high\ group produced \exact\ answers compared to 34.59\% in the \low\ group. 
Additionally, \low\ had a median \error\ of 0.176 and \high\ had a median \error\ of 0. Consistent with the regression analysis, we show a statistically significant difference between the \low\ and \high\ groups when we compared \exact\ $(\chi^2(1) = 6.4762, p = 0.0109, d= 0.3980)$ and \error\ (Kruskal-Wallis, $H(1) = 6.8535, p =  0.0088,  \eta^2(H) = 0.0348$).
Together, these results support \textbf{H2}, showing \textit{suggestive evidence that participants' working memory mediated Bayesian reasoning accuracy}.

\subsubsection{\textbf{Working Memory Capacity \& Presentation Formats}}

\lb{}
In light of our previous finding that successful reasoning might depend on cognitive resources, we conducted further analysis to examine the effect of presentation format on reasoning accuracy within the \low\ and \high\ groups. Specifically, we ran separate 3-way proportion tests to compare the frequencies of \exact\ answers and found no difference between presentation formats for both \low\ $(\chi^2(2) = 4.8401, p =  0.0889)$ and \high\ $(\chi^2(2) = 0.6504, p =  0.7224)$ groups.
Further, a Kruskal-Wallis test comparing \error\ for \textit{text}, \vis, and \vistext\ within the \low\ group revealed a statistically significant difference between the three presentation formats ($H(2) = 10.086, p =   0.006453,  \eta^2(H)=0.0817 $).  We ran pairwise Mann-Whitney Wilcoxon tests with an adjusted alpha $(\alpha=0.0167)$  and found significant differences in \error\ between \textit{text} and \vis\ ( $W = 874, p =  .0017, \eta^2(H)=0.1291$), but failed to reject the null hypothesis for the \textit{text} \& \vistext\ and \vis\ \& \vistext\ comparisons. Examining the \high\ group, a Kruskal-Wallis test found no overall significant differences between presentation formats ($H(2) = 1.7301, p = 0.4210$). These analyses suggest that \textit{presentation choices can impact users with low working memory capacity}, with \textit{text} eliciting significantly higher error rates compared to \vis. However, the \textit{high working memory capacity participants were less impacted by the format they used}.

Our final analysis here investigates how \low\ and \high\ groups performed within each presentation condition. Our analysis revealed that the \low\ and \high\ groups had similar proportions of \exact\ $(\chi^2(1) = 1.2013, p =  0.2731)$ answers and \error\ ($W=428.5, p =  0.4381$) rates when reasoning with \vis. The two working memory groups also did not differ in \exact\ $(\chi^2(1) = 1.5875, p =  0.2077)$ and \error\ when using \vistext\ ($W=230, p =  0.1000$). However, we observed a statistically significant difference in \exact $(\chi^2(1) = 5.889, p =  0.01524, d=0.6997)$ and \error\ with the \textit{text} condition ($W=235.5, p =  0.02481, \eta^2(H)=0.0776$). Thus, \textit{text} is marginally more likely to elicit a deviation in accuracy between \low\ and \high\ compared to \vis\ or \vistext.

\subsubsection{\textbf{Dual Task: Reasoning Under Divided Attention}}

\lb{}
In \textbf{H1}, we posit that if we can experimentally manipulate executive capacity by adding a secondary task, we will incur a decline in performance, known as the \textbf{dual-task cost}. As a result, formats that require high cognitive resources will have a significant dual-task cost.

\paragraph{\exact} We observed a near-identical proportion of \exact\ answers for the \single\ and \dual\ conditions. Participants in the \dual\ condition produced the \exact\ answer 41.86\% of the time, compared to 41.17\% in the \single\ condition.  We compared the proportion of \exact\ answers in the \single\ and \dual\ task and found no overall significant differences $(\chi^2(2) = 0.0141, p = 0.9054)$. The analysis revealed 34\% of \exact\ answers for \txt, 57.14\% for \vis, and 38.64\% for \vistext. A 3-sample proportion test found no significant difference in \exact\ between the presentation formats in the \dual\ group $(\chi^2(2) = 4.816, p = .09)$.\textit{Thus, manipulating load had no significant effect on our participants' \exact\ responses.}

\paragraph{\bias} A Kruskal-Wallis test found no significant difference in \bias\ between presentation formats in the \dual\ group ($H(2) = 4.44, p = 0.1084$).
We compared overall \bias\ for the \single\ and \dual\ conditions using a Mann-Whitney Wilcoxon test and found no significant difference between the two conditions ($W= 11130, p= 0.8175$).

\paragraph{\error} Finally, we also observed an identical overall median \error\ of 0.097 for both the \single\ and \dual\ task. An overall comparison with a Mann-Whitney Wilcoxon test found no significant difference in \error\ between \single\ and \dual\ ($W= 11232, p= 0.709$). The median \error\ was 0.27 for \txt, 0 for \vis, and 0.097 for \vistext\ in the \dual\ task. Similar to the \single\ condition, an omnibus Kruskal-Wallis test revealed an overall effect of presentation format on \error\ in the \dual\ condition ($H(2) = 7.7344, p = .0207, \eta^2(H)= 0.0455$). Follow-up Mann-Whitney Wilcoxon tests with an adjusted alpha ($(\alpha= .0167)$ revealed significant differences in \error\ between \textit{text} and \vis\ ($W = 1162.5, p = .0073 , \eta^2(H)=0.0747$). We found no significant difference between \textit{text} and \vistext\ ($W = 1277.5, p =0.1674$) and \vis\ and \vistext\ ($W = 612.5, p =0.1002$).

\begin{figure*}[h!]
    \centering
    \includegraphics[width=1\textwidth]{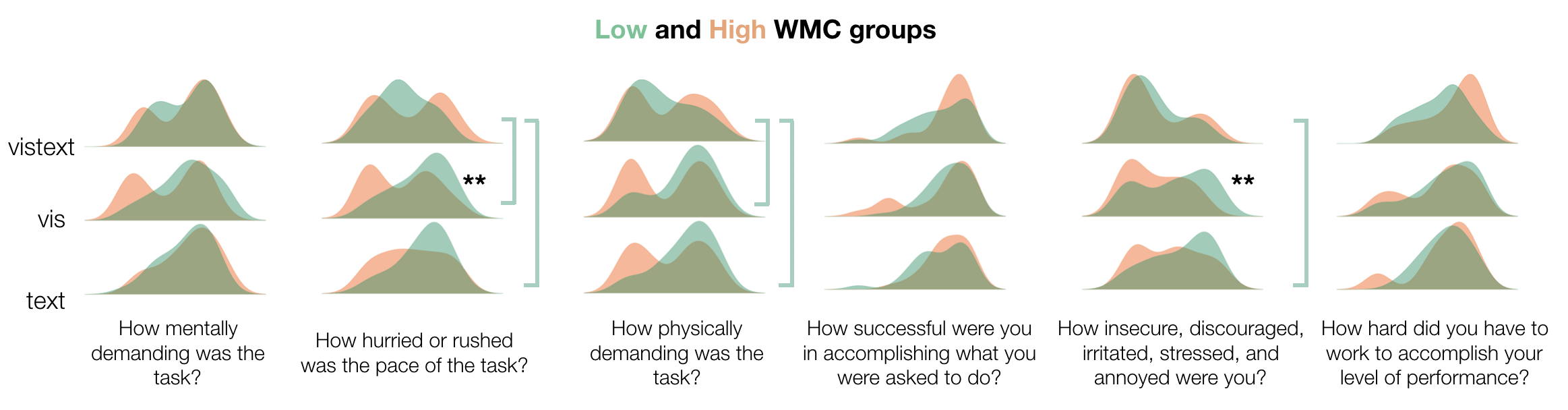}
    \vspace{-.5em}
    \caption{Distribution of \tlx\ scores in the \single\ task for the \low and \high \wmc\ groups. ** indicates a statistically significant difference between working memory groups $(\alpha<0.05)$ and $]$ indicates differences across formats for the corresponding \wmc\ group $(\alpha<0.0167)$.}
    \label{fig:nasatlx_wmc}
\end{figure*}

\textbf{Considering differences in working memory capacity.} In section \ref{sssec:workingmemFindingsExp1}, we showed evidence that working memory capacity impacts Bayesian reasoning. Here, we examine the difference in performance between the \single\ and \dual\ conditions by taking into account individual differences in working memory capacity. For individuals in the \high\ group, we found no significant difference between those in the \single\ and \dual\ task conditions when examining \textsc{exact} $(\chi^2(2) =0.4258, p = 0.5141)$  and \textsc{error} ($W = 3104, p = 0.3264$). Similarly, we found no measurable difference between the \single\ and \dual\ conditions for participants in the \low\ groups when examining \textsc{exact}$(\chi^2(2) =0.0701, p = 0.7912)$  and \textsc{error} ($W = 2467, p = 0.5253$)
Taken together, the secondary task did not elicit the expected results and the evidence for \textbf{H1} is inconclusive. %

Although in section \ref{ssec:exp1attrition} we found no significant difference in attrition rate between the \single\ and \dual\ tasks, we conducted a Kruskal-Wallis test to investigate whether the distribution of \ospan scores varied between the two tasks after all data quality exclusions. We found a significant difference in \ospan scores between \single\ and \dual\ ($W = 13688, p =  0.0002, \eta^2(H)=0.0422$), with higher \ospan scores in the \dual\ condition. This could be due to selective attrition or bias in our sample. Participants in the \dual\ condition had significantly higher \ospan scores compared to the \single\ task. This could also explain why we did not observe a significant decline in performance between the two tasks. We will consider this confounding factor in our interpretation of Experiment 1's results.

\subsubsection{\textbf{\tlx\ Self-Reported Effort}}
\lb{}
When looking at self-reported effort in the \single\ task, we found an overall significant difference in perceived \textbf{frustration} across presentation formats (Kruskal-Wallis, $H(2) = 11.72, p =  0.003, \eta^2(H)=0.0582$). We conducted separate Mann-Whitney Wilcoxon tests with an adjusted alpha $\alpha=0.0167$ for pairwise comparisons that revealed a significant difference in \textbf{frustration} between \vistext\ and \textit{text} ($W = 1918.5, p= 0.0005, \eta^2(H)=0.1078$). 

Since working memory capacity is likely to affect reported NASA-TLX scores, we observed differences between presentation formats for each working memory group separately. We found no significant difference between presentation formats across any of the \tlx\ subscales in the \high working memory group. Within the \low\ group, we conducted separate Kruskal-Wallis tests and found significant differences in accuracy between presentation formats in the following:

\begin{itemize}
    \item \textit{temporal demand:} ($H(2) = 10.305, p =  0.006, \eta^2(H)=0.0839$)
    \item \textit{physical demand:} ($H(2) = 8.95, p =  0.0114, \eta^2(H)=-0.0045$)
    \item \textit{frustration:} ($H(2) = 10.825, p =  0.004, \eta^2(H)=0.089$)
\end{itemize}  

\vspace{1em}
\noindent
As a follow-up, we conducted Mann-Whitney Wilcoxon tests with an adjusted alpha $(\alpha=0.0167)$ within the \low\ group and found significant differences between \vistext\ and \textit{text} in the following:

\begin{itemize}
    \item \textit{temporal demand} ($W = 644, p= 0.004, \eta^2(H)=0.1262$),
    \item \textit{physical demand } ($W = 638, p = 0.005,\eta^2(H)=0.1175 $)
    \item \textit{frustration} ($W = 672.5, p = 0.0009, \eta^2(H)=0.1712$)
\end{itemize}  

\vspace{1em}
\noindent
Within the \low\ group, we also found differences between \textit{vis} and \textit{vistext}  in the following:

\begin{itemize}
    \item \textit{temporal demand} ($W = 894.5, p= 0.006,  \eta^2(H)=0.0904$),
    \item \textit{physical demand} ($W = 878, p= 0.011,  \eta^2(H)=0.08296$)
\end{itemize}  

\vspace{1em}
\noindent
We investigated differences in reported \mbox{\tlx} scores across \high and \low groups for each presentation format. In the \vis\ condition, we found differences in the following:

\begin{itemize}
    \item \textit{temporal demand} ($W = 307, p= 0.01525, \eta^2(H)= 0.0673$)
    \item \textit{frustration} ($W = 332.5, p = 0.0379,  \eta^2(H)=0.0525$)
\end{itemize}

Finally, we found no differences in reported scores between working memory groups in the \textit{text} and the \vis\ conditions.

\subsubsection{\textbf{Additional Analyses}}

\paragraph{Spatial Ability}
We conducted a generalized linear model with a logit link and found that spatial ability score had a significant impact on \exact\ $(z(298)=4.670, p=3.00e-06)$. We also examined the effect of spatial ability score on \error\ through a generalized linear model and found significant effects $(t(298)=-4.003, p= 7.91e-05)$. These findings replicate prior work showing that spatial ability mediates Bayesian reasoning\cite{ottley2016improving,mosca2021does}. 

\paragraph{Completion Time}

Kruskal-Wallis tests revealed no significant effect of presentation format ($ H(2)= 1.2454, p= 0.5365$) or load condition ($ H(2)= 2.03, p= 0.1546$) on the completion time of the Bayesian task. Moreover, we found no significant difference in completion time between the \low\ and \high\ working memory groups ($ H(2)= 0.0797, p= 0.7778$).

\paragraph{Cognitive Reflection Test}
Our \crt\ results largely replicated the \ospan\ findings. We found an overall significant impact of \crt\ score on \exact\ $(\chi^2 (3)= 20.502, p=0.0001, \eta^2(H)=0.5246)$. Further, the Kruskal-Wallis test shows a statistically significant effect of \crt\ on  \bias\ ($H(3)= 20.566, p=0.0001, \eta^2(H)=0.0595)$ and \error\ ($H(3)= 24.986, p=1.555e-05, \eta^2(H)=0.0745)$, showing evidence that \textit{individuals with a higher \crt\ score were significantly more likely to enter the exact answers and made smaller reasoning errors.}

\section{Experiment 2: Mixed Design Study}

Experiment 1 used a between-subject design to control for learning effects and ensured consistency by comparing responses to the same Bayesian problem. However, we found no significant effect of the dual task on accuracy. This could be due to the differences in working memory capacity between the two groups, or high individual variability due to the study design.
We conducted a second mixed design study\footnote{Link to Experiment 2 surveys, data, and analyses: \url{https://bit.ly/3LfgXCZ}} to 1) control for individual variability in the single and dual tasks and 2) test whether the lack of replication is due to population or methodological differences.

We made the following changes to the experiment design to reduce the overall difficulty of the task and better control for individual variability. 
\begin{itemize}
    \item \textbf{Improve Study Preparation with a Practice Round}: We added a pre-study trial to familiarize participants with the task and study structure. Participants saw and attempted a sample Bayesian reasoning task before continuing to the main task.
    \item \textbf{Control Individual Variability}: We used a mixed factorial design with the load condition (\single, \dual) as a within-subject factor and presentation format (\textit{vis}, \textit{text}, \textit{vistext}) as a between-subject factor. 
    \item \textbf{Remove CRT Test}: We removed the Cognitive Reflection Test from the survey as we found in Experiment 1 that it is positively correlated with the OSPAN test $(r(297) = 0.27, p= 2.051e^{-06})$, which is more widely recognized \mbox{\cite{engle1999individual,conway2005working,oswald2015development,unsworth2005automated}}. This shortened the survey.
\end{itemize}

\begin{table}[h]
  \caption{Scenarios use in the Bayesian task in Experiment 2}
  \label{tab:commands}
  \begin{tabular}{cp{6cm}}
    \toprule
    Scenario & Description\\
    \midrule
    \textit{cab} & There is a total of 100 witnesses to the car accident. Out of the 100 witnesses, 15 claimed that the car which caused the accident was a cab. Out of these 15 witnesses, 12 claimed the car was blue and 3 claimed the car was green. On the other hand, 85 witnesses claimed that the car which caused the accident was not a cab. Out of these 85 witnesses, 3 claimed the car was blue and 82 claimed the car was green.\\ \\
    \textit{class} & There is a total of 100 college freshmen in the population. Out of these 100 freshmen, 30 are enrolled in an introductory entrepreneurship course. Out of these 30 freshmen, 20 plan on going into business after graduation, and 10 do not. On the other hand, 70 freshmen are not enrolled in an introductory entrepreneurship course. Out of these 70 freshmen, 10 plan on going into business after graduation, and 60 do not. \\
    \bottomrule
  \end{tabular}
\end{table}

\subsection{Task \& Procedures}

Our tasks were similar to Experiment 1 except that in the Bayesian survey, the users conducted both a single and dual task in no particular order where the Bayesian problems were presented using two scenarios, a \textit{cab} and a \textit{class} scenario. After solving the Bayesian problems, users completed a NASA-TLX and a spatial ability test. In this experiment, we did not conduct the Cognitive Reflection Test. Similarly to Experiment 1, users also completed an OSPAN test. 

\textbf{Practice Round}. In the practice round, users practiced the dot pattern recall task, the \single Bayesian task, as well as both tasks together as part of the \dual condition.

\textbf{Payment.} Participants received a base pay of \$2 and could win a total bonus of up to \$2.5, comprising of \$0.5 for each correct Bayesian question and \$0.5 for a correctly reproduced dot pattern. Participants received an average bonus of \$1.51 and completed the Bayesian and OSPAN surveys in an average time of 26.8 minutes. 

\subsection{Experimental Design}
    Similarly to Experiment 1, we assigned each user randomly to one of three presentation conditions (\textit{vis}, \textit{text} or \textit{vistext}), making the comparison of presentation between subjects. Each user completed both the \single\ and \dual\ tasks and saw either the \textit{cab} or \textit{car} scenario, making load condition a within-subject condition.

\subsection{Presentation Conditions}

Our presentation formats remained a between-factor condition and were the same as Experiment 1: \textit{text}, \vis, and \vistext. We utilized the \textit{disease} scenario for the pre-task tutorial, and each participant saw two Bayesian problems narrating two different scenarios: \textit{cab} and \textit{class}~\cite{micallef2012assessing,gigerenzer1995improve,ottley2019curious}. The \textit{cab} scenario involves eye-witness testimonies of a hit-and-run scenario, while the \textit{class} scenario presents the career prospects of college students. We randomly assigned one scenario to the \single\ task and the other to the \dual\ condition, and the order of the conditions was counterbalanced.

\subsection{Participants}
As per our pre-registration \footnote{Link to Experiment 2 pre-registration: \url{https://bit.ly/3qIzJJu}}, we conducted a power analysis based on a three-way mixed ANOVA and determined the ideal sample size to be 168. We recruited 240 participants via Amazon's Mechanical Turk to account for a 30-40\% exclusion rate. Participants were English-speaking from the United States and had a HIT acceptance rate of 100\%. After excluding 88 participants based on the same pre-registered criteria determined in Experiment 1 (see section \ref{ssec:participantsExp1}), 152 participants remained (\textit{text}= 46, \textit{vis}=55, \textit{vistext}=51).

\subsubsection{Attrition Rate}

Using the same methodology as Experiment 1, we conducted an attrition rate analysis for Experiment 2. Table~\ref{tab:attritionExp2} shows the condition-wise dropout rate, showing participants who first saw the \dual task or the \single task. We found no significant difference in dropout rate between the two conditions $(\chi^2(2) = 1.6824, p=0.1946, d=0.1293)$.
When looking at performance in the \ospan\ test after exclusions, we found no significant difference in scores. This suggests that the population who completed the experiment was consistent across both conditions ($H(2) = 0.34157, p =  0.5589, \eta^2(H)=-0.0048$).

\begin{table}[h]
  \caption{Experiment 2 condition-wise dropout rates for the Bayesian Task}
  \label{tab:attritionExp2}
  \begin{tabular}{p{6cm}c}
    \toprule
    Task Order & Dropout rates\\
    \midrule
    \single, \dual: Participants saw the dual task followed by the single task & 11.23\% \\
    \dual, \single: Participants saw the single task followed by the dual task & 15.67\% \\
    \bottomrule
  \end{tabular}
\end{table}

\subsection{Results}

In this experiment, our aim is to uncover differences in dual-task costs elicited by each presentation format through a mixed-design study. First, we establish a baseline for accuracy in the single task and compare our findings to Experiment 1. Then, we examine and compare the decline in performance elicited by the dual task \textbf{(dual-task cost)} between presentation formats. 

\begin{figure*}[h]
     \centering
         \centering
         \includegraphics[width=0.75\linewidth]{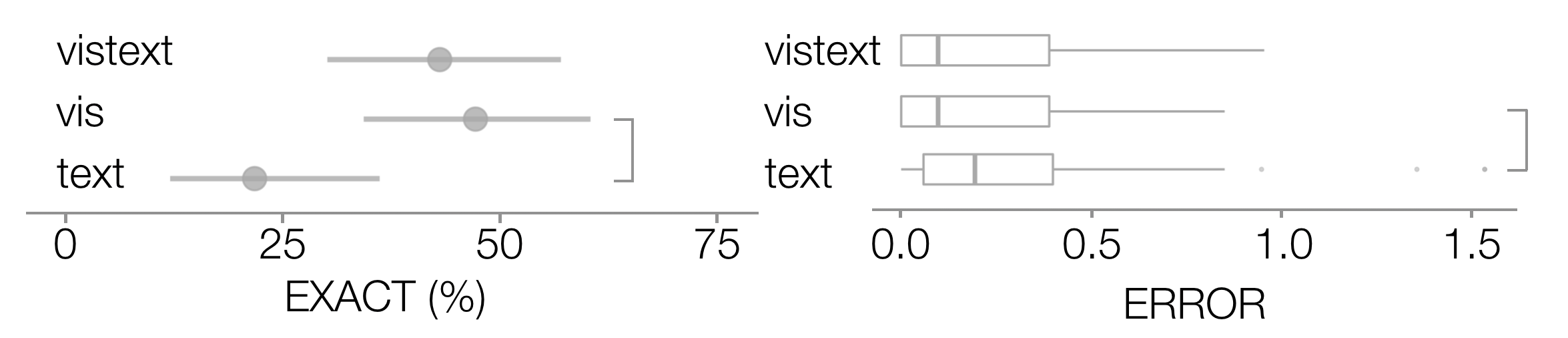}
        \caption{Experiment 2. \single\ task \exact, and \error\ across presentation formats. {\Large{]}} indicates a significant difference between the two formats ($\alpha=0.0167$).}
        \label{fig:e2_overview}
\end{figure*}

\subsubsection{Single Task}

\paragraph{\exact}  Overall, 38.2\% of participants correctly answered both Bayesian questions in the \single task, with \txt, \vis, and \vistext\ leading 21.7\%, 47.3\% and 42.3\% \exact\ answers respectively. Contrarily to Experiment 1, our analysis shows a significant effect of presentation format on \exact\ $(\chi^2(2) = 7.58, p = .023, d=0.4584)$. Follow-up pairwise 2-sample proportion tests with an adjusted alpha $(\alpha=0.0167)$ revealed a significant difference in \exact\ between \vis\ and \textit{text}  $(\chi^2(2) = 7.1195, p = 0.0076, d=0.5537)$.

\paragraph{\bias} We found no significant difference in \bias\ between the presentation formats (Kruskal-Wallis, $H(2) = 1.0826, p = 0.582, \eta^2(H)=-0.0063)$.

\paragraph{\error} The median \error\ was 0.097 overall and 0.097 in the \vis\ condition, 0.194 in the \textit{text} condition and 0.076 in the \vistext\ condition. A Kruskal-Wallis non-parametric test found a significant difference in \error\ between the presentation formats ($H(2) = 7.61, p = 0.0223, \eta^2(H)=0.0376$). 
Post-hoc Mann-Whitney Wilcoxon tests with an adjusted alpha $(\alpha=0.0167)$ revealed a significant difference between \vis\ and \textit{text} ($W = 1619.5, p= 0.01329, \eta^2(H)=0.05182$). 
Overall, the general trends are in line with Experiment 1 and demonstrate that \textit{participants were the least accurate with text compared to visualization}.  However, the differences are more pronounced in Experiment~2.  

Prior work has shown that different Bayesian scenarios can have a different impact on accuracy \cite{micallef2012assessing}. We found no significant difference in accuracy between the \textit{class} and \textit{cab} scenarios when looking at \exact $(\chi^2(2) = 0.0251, p = 0.8741, d=0.0257)$, \bias ($W = 2870.5, p= 0.9727, \eta^2(H)=-0.0067$) or \error ($W = 2952.5, p= 0.7844, \eta^2(H)=-0.00616$).

\subsubsection{Single vs Dual Task}

\begin{figure}[b]
         \centering
         \includegraphics[width=\linewidth]{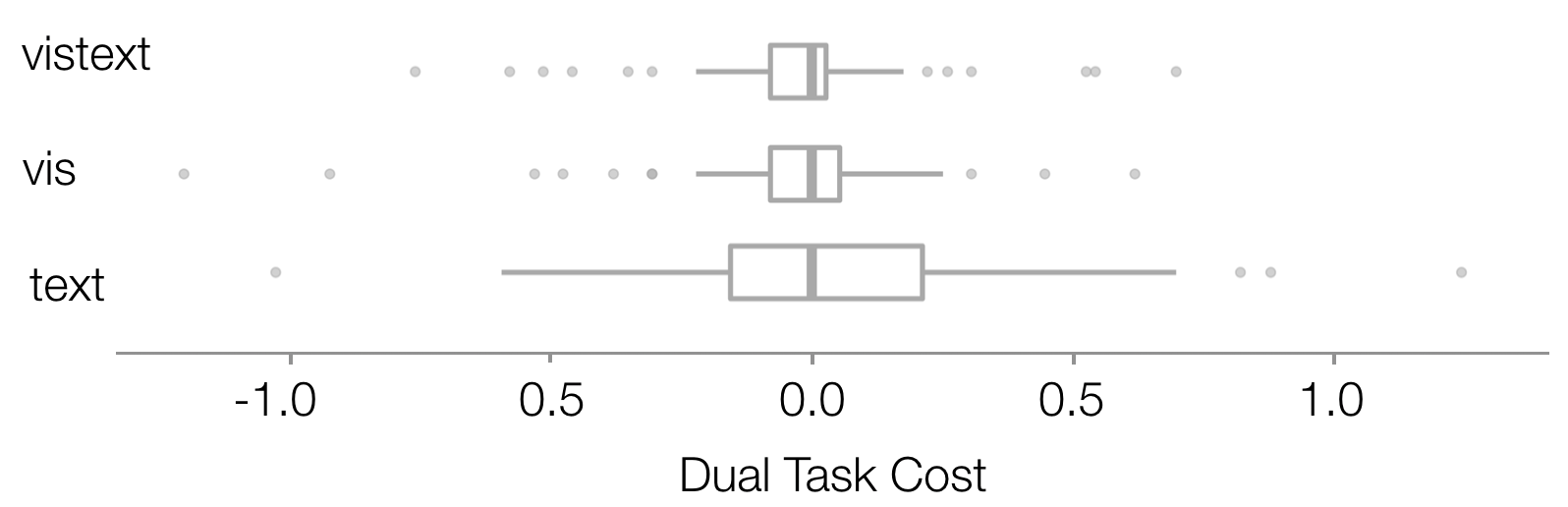}
         \label{fig:error_wmc_noldsoad}
         \vspace{-1.5em}
        \caption{Dual-task cost across presentation formats.}
        \label{fig:exp1_exact}
\end{figure}

\paragraph{\textsc{Dual-task}} We found that in the \dual task, the mean number of dots recalled was 3.56 $(\sigma=0.77)$. 40.8\% of participants correctly answers both Bayesian questions, with 26.1\% of \exact answers for \textit{text}, 41.8\% for \vis\ and 51.9\% for \vistext. Overall differences in \exact\ between presentation formats were significantly different $(\chi^2(2) = 6.8197, p = 0.0331, d=0.4335)$. Follow-up pairwise comparisons revealed a significant difference in \exact\ between \textit{text} and \vistext\ $(\chi^2(2) = 6.8003, p =0.0091, d=0.5461)$, \textit{suggesting that the combination of visualization and text leads to fewer errors than text alone under divided attention}. We found no significant difference in \bias\ (Kruskal-Wallis, $H(2) = 2.3795, p = 0.3043$) or \error\ (Kruskal-Wallis, $H(2) = 4.451, p = 0.108$) between presentation formats.

\paragraph{\textsc{Dual-task cost}} For each participant, we observed the decline in performance in the dual task compared to the single task by computing the difference in \error, a measure known as the \textbf{dual-task cost}. By comparing dual-task costs across presentation formats, we can infer differences in cognitive load. We conducted a Kruskal-Wallis test and found no overall difference in dual-task costs between presentation formats ($H(2) = 1.0314, p = 0.5971, \eta^2(H)= -0.0065$).

\paragraph{\textsc{Calibrating Dual-task cost}} In section \ref{sssec:workingmemFindingsExp1}, we showed evidence that working memory capacity impacts Bayesian reasoning. To this end, we observe the effect of dual-task cost for \high and \low working memory capacity groups. We found that dual-task costs were not significantly different between presentation formats within the \high (Kruskal-Wallis, $H(2) = 1.4324, p = .4886, \eta^2(H)=-0.0089$) or \low (Kruskal-Wallis, $H(2) = .14215, p = .9314, \eta^2(H)=-0.0226$) group. \textit{Therefore, we conclude that even when considering individual differences in working memory capacity, the effect of the dual-task was consistent across presentation formats.} 

\section{Discussion}

Our work leveraged cognitive theory to understand the conflicting findings about the effect of combining text and visualization in the context of Bayesian communication. Analyzing general trends in accuracy alone seldom paints the complete picture in an evaluation study, as there is ample research on the impact of individual differences on Bayesian reasoning and beyond~\cite{ottley2016improving,mosca2021does, yin2020bayesian,castro2019cognitive}. Our results suggest that combining visualization and text does not increase cognitive load and in some cases improves subjective workload. We present our main takeaways from these studies.

We analyzed accuracy to compare our findings to the prior work and to provide context for our cognitive load results. At a high level, our produced  results are similar to prior studies in the visualization community ~\cite{mosca2021does,ottley2019curious,ottley2016improving,micallef2012assessing}.
\textit{Presentation format alone had little to no effect on Bayesian reasoning, but the inclusion of visualization improved Bayesian reasoning}. In Experiment 1, we found that users' proportion of correct answers in the baseline \noload condition was not significantly different across the three formats, replicating Ottley et al.'s findings~\cite{ottley2016improving}. While user error rates were significantly lower using visualization-only compared to text-only, the effect size was small for the statistical test. In Experiment 2, we saw a significantly greater proportion of correct answers with the combined presentation format than with text alone, with a small effect. Still, although our accuracy analysis does a good job of uncovering differences, it does not explain the phenomena. 

\subsection{Implications of Cognitive Load}

We leveraged three different but complementary techniques for evaluating cognitive load: a working memory capacity test, self-reported effort, and a dual-task. Our investigations into working memory capacity were influenced by prior work, primarily focusing on text-only formats, and showed a positive correlation between working memory capacity and reasoning performance~\cite{lesage2013evolutionary, yin2020bayesian}. In our work, we found that the effect of working memory capacity held, with high working memory individuals generally outperforming their low working memory counterparts. This effect was especially salient in the text-only condition. These findings help contextualize our findings on Bayesian task accuracy, which suggest that visualization and multimedia formats may be superior to text-only. 

We also saw that participants with low working memory capacity performed better when using visualization alone than text alone. In line with Castro et al.'s work~\cite{castro2021}, this difference in performance in the low working memory group is indirect evidence that text-only elicited more cognitive load than visualization-only. Expanding this argument, we can deduce that the combination of text and visualization did not elicit more cognitive load than visualization-only. Given the prior findings that removing numbers from the text in the combined presentation positively affects reasoning performance\cite{micallef2012assessing}, we expected to find evidence that combining text and visualization increases cognitive load, but our data does not support this notion. These findings have practical implications for visualization recommendation and accessibility. Visualizations can benefit populations with lower cognitive abilities and be beneficial in situations of high cognitive burden.

Notably, in Experiment 1, participants with low working memory capacity reported experiencing significantly lower frustration and temporal demand when using the combination format compared to text alone. This finding further supports the use of the multimedia format. However, it is noteworthy that Experiment 1 also showed no significant difference in the accuracy rates between the combination format and text for individuals with low working memory, highlighting the deficiency of analyzing accuracy alone. In general, our findings somewhat support the notion of a multimedia effect. In particular, there may be some benefit to having both text and visualization available to facilitate reasoning, especially for people with low working memory capacity. One potential explanation might be that visualization allows the viewer to offload items from memory, but the text is familiar and easy to process. This hypothesis corroborates the results of prior work that captured eye-gaze data as people solved Bayesian tasks~\cite{ottley2019curious}. Their results suggest that visualization makes it easy to identify relevant information, but the text may be easier to process compared to the visual format~\cite{ottley2019curious}.  Another plausible explanation for our results is that participants with low working memory might prefer the flexibility of the combined format, which enables them to choose the format that best aligns with their mental model or preference. Further investigation is needed to better understand this phenomenon.

\subsection{On The Failure of the Dual-Task Paradigm}

The dual-task paradigm did not reveal differences in cognitive load across formats, even when accounting for individual differences in working memory capacity. Specifically, asking participants to hold a dot pattern in memory did not influence their reasoning performance. We hypothesized in Experiment 1 that this effect could be due to individual variability in the between-subject design. However, the within-subject Experiment 2 revealed similar findings, which contradicts H3 and prior work \cite{lesage2013evolutionary} possibly due to differences in experimental design. For example, Lesage et al.\cite{lesage2013evolutionary} performed a laboratory experiment with 179 first-year psychology students who participated in the previous study for course credits. Our study used a more diverse crowdsourced study population. Another possible explanation is that the secondary task was too easy, or our study participants may have written down the pattern instead of holding it in memory. Alternatively, the observed disparity may also be due to differences in the demographic makeup of our study populations. 

Several researchers have developed guidelines for choosing an adequate secondary task, which includes considerations for task difficulty and similarity \cite{sanders1987mccormick, padilla2019toward}. However, it can be challenging to strike the perfect balance between the primary and secondary task as the latter has to be hard enough to increase cognitive load but not to the point of cognitive overload. 
Although the exact reason for the failed replication is unknown, we encourage researchers to consider modifying the dual-task methodology when conducting crowdsourced evaluations. One alternative study design could be to calibrate the secondary task's difficulty based on participants' abilities. For example, Castro et al.~\cite{castro2019cognitive} used calibration in a study investigating the impact of divided attention on driving. Their participants performed a pre-test to identify the level of difficulty that elicited a 75\% accuracy on the secondary task. For our choice of secondary task, one option would be to calibrate the size of the dot pattern to memorize based on participants' performance. Alternatively, Borgo et al.'s study on the impact of visual embellishment on engagement and working memory used a word selection secondary task~\cite{borgo2012empirical}, where users identified fruits among a crawling list of words. Researchers could calibrate the secondary task by tailoring the crawl speed of the words to each participant. Further, Borgo et al.'s~\cite{borgo2012empirical} dual-task setup would be less susceptible to violations of the study tasks since it does not involve a recall task. 

\section{Conclusion}
Our work expands the understanding of the relationship between working memory capacity and Bayesian reasoning by examining and comparing three presentation formats. By examining more granular accuracy measures, we showed that visualization-only and combination formats lead to less error in Bayesian reasoning than text-only formats. Moreover, we showed that working memory capacity mediates Bayesian reasoning accuracy, particularly in the text format. Finally, we showed that users with low working memory capacity are more accurate when using visualization alone compared to text alone. We discuss how these findings can impact visualization design guidelines, especially for low working memory capacity users. To this end, we argue for more diversified evaluation metrics and encourage the visualization community to leverage and apply existing research in cognitive science and related fields to better understand how people perceive and reason with visualizations.

\begin{acks}
The authors wish to thank Lace Padilla for her insights into the use of cognitive methods and Shayan Monadjemi for his valuable feedback on the manuscript. This material is based upon work supported by the National Science Foundation under grant number 2142977. 
\end{acks}

\bibliographystyle{ACM-Reference-Format}
\bibliography{MAIN-publication}

\end{document}